\documentclass[12pt]{article}
\usepackage[a4paper,margin=2.5cm]{geometry}
\usepackage{lipsum}
\usepackage{authblk}

\usepackage{amsmath,amssymb,bm,graphicx}
\usepackage{amsthm}
\usepackage{multirow}
\usepackage{dcolumn}
\usepackage{latexsym}
\usepackage{hyperref}
\usepackage[normalem]{ulem}
\usepackage{physics}
\usepackage{mathtools}

\usepackage{cite}
\usepackage{comment}
\usepackage{titlesec} 
\usepackage{chngcntr} 
\usepackage{orcidlink}

\newcommand{\Exp}[1]{\langle #1 \rangle}

\definecolor{darkgreen}{rgb}{0.0,0.45,0.0}
\newcommand{\GRN}[1]{{\color{black}#1}}
\newcommand{\GEN}[1]{{\color{black}#1}}

\newcommand{\JAEHA}[1]{{\color{black}#1}}
\newcommand{\RED}[1]{{\color{black}#1}}

\newcommand{\I}{\mathop{\mathbb{I}}\nolimits}

\newcommand{\CA}{\mathop{\mathbb{C}}\nolimits}

\newcommand{\Md}{M_d(\mathbb{C})}

\newcommand{\defeq}{\coloneqq}

\newtheorem{Theorem}{Theorem}
\newtheorem{Proposition}{Proposition}

\newtheorem{Lemma}{Lemma}
\newtheorem{Remark}{Remark}

\newtheorem*{Proof*}{Proof}

\newcommand{\lm}{\lambda_{\rm 1}}
\newcommand{\lSm}{\lambda_{\rm 2}}
\newcommand{\rhonorm}[1]{\| #1 \|_\rho}

\newcommand{\zExp}[1]{{#1}_{\rm shift}}

\begin{document}

\title{Beyond Robertson--Schr\"odinger: A General Uncertainty Relation Unveiling Hidden Noncommutative Trade-offs}

\author[1]{Gen Kimura\,\orcidlink{0000-0003-4288-2024}\thanks{\href{gen.kimura.quant@gmail.com}{gen.kimura.quant@gmail.com}}}
\author[2]{Aina Mayumi\,\thanks{\href{a.mayumi1441@gmail.com}{a.mayumi1441@gmail.com}}}
\author[3]{Hiromichi Ohno\,\orcidlink{0000-0001-5498-3311}\thanks{\href{h_ohno@shinshu-u.ac.jp}{h\_ohno@shinshu-u.ac.jp}}}
\author[4]{Jaeha Lee}
\author[5]{Dariusz Chru\'sci\'nski \orcidlink{0000-0002-6582-6730}}

\affil[1]{Graduate School of Information Sciences, Tohoku University, Aoba-ku, Sendai 980-8579, Japan}
\affil[2]{College of Systems Engineering and Science, Shibaura Institute of Technology, Saitama 330-8570, Japan}
\affil[3]{Department of Mathematics, Faculty of Engineering, Shinshu University, 4-17-1 Wakasato, Nagano 380-8553, Japan}
\affil[4]{Institute of Industrial Science, The University of Tokyo, Chiba 277-8574, Japan}
\affil[5]{Institute of Physics, Faculty of Physics, Astronomy and Informatics, Nicolaus Copernicus University, Grudziadzka 5, 87-100 Torun, Poland}

\maketitle

\begin{abstract}
We report a universal improvement to the standard Robertson--Schr\"odinger uncertainty relation. 
Our result shows that the Robertson--Schr\"odinger lower bound can be supplemented by a new noncommutativity-induced term. 
This term represents a previously overlooked quantum contribution and becomes more pronounced as the state becomes more mixed.
Moreover, it is expressed as the expectation value of a positive observable, namely the squared modulus of the commutator, and therefore preserves the direct, experimentally accessible character of the Robertson--Schr\"odinger relation.
For two-level quantum systems, our relation becomes an \emph{exact equality} for \emph{any} state and \emph{any} pair of observables, thereby ensuring the tightness of the bound in the strongest possible sense.
The relation also yields, as a corollary, a complete proof of a general uncertainty bound that had previously been supported only by numerical evidence.
\end{abstract}

\section*{Introduction}
\paragraph*{Early Developments.}
The uncertainty principle, exposed by Heisenberg in 1927 \cite{Heisenberg1927}, is widely regarded as one of the hallmarks of quantum theory, signalling its departure from classical intuition.  Among its profound implications is the indeterminacy of quantum states, which is most typically exemplified by the fact that the position-momentum pair never admits a determinate joint description in the microscopic world.

Following Heisenberg's seminal yet arguably nebulous exposition of the notion of uncertainty (or `indeterminateness' \cite{Heisenberg1930}), where he heuristically communicated the idea by examining the fundamental limitation on the joint description of the position and momentum of a particle, its first mathematically unambiguous formulation was soon after given by Kennard~\cite{Kennard1927};  by interpreting the uncertainties of the canonically conjugate pair as their standard deviations, their product is found to be never less than half of the reduced Planck constant.  Building on top of Weyl's modernized proof~\cite{Weyl1928} inspired by Pauli's remark, Robertson~\cite{Robertson1929} further generalized the relation to arbitrary pairs of observable functions on the coordinate space.  Their contributions eventually lead to what is now widely referred to as the Robertson uncertainty relation
\begin{equation}\label{eq:RUR}
V_\rho(A) V_\rho(B) \ge \frac{1}{4}|\langle [A,B]\rangle_\rho |^2
\end{equation}
with $\langle X \rangle_\rho \defeq \Tr [X\rho]$ and $V_\rho(X) \defeq \Tr[(X-\langle X \rangle_\rho)^2\rho]$ respectively denoting the expectation value and the variance of an observable $X$ under a quantum state $\rho$. 
\GEN{The product of the uncertainties of a pair of observables is bounded from below by a term involving their commutator, $[A,B] \defeq AB-BA$.
Robertson's relation may therefore be viewed as the first general formulation that made explicit the link between uncertainty trade-offs and the noncommutativity of observables.}

\paragraph*{Landscape of modern refinements.}

\GEN{In modern terminology, the Robertson relation is usually classified as a preparation uncertainty relation. 
It concerns the statistical uncertainty of measurement outcomes inherent in a prepared quantum state, rather than the additional error or disturbance caused by a particular measurement process. 
This distinction separates it from Heisenberg's earlier discussions, including the famous gamma-ray microscope Gedankenexperiment, where measurement limitations and disturbance played a central role. 
Uncertainty relations that explicitly address measurement error and disturbance in the spirit of Heisenberg's original discussions are usually referred to as measurement uncertainty relations, and have been extensively developed in various formulations (see, e.g., \cite{Arthurs1988,OZAWA2004350,Werner2004,BUSCH2007155,Miyadera2008,Watanabe2011,Branciard2013,Lee2024}).}

Meanwhile, the preparation uncertainty relations, brought forth by Kennard and Robertson, have also seen various improvements and modifications. One of the earliest contributions was Schr{\"o}dinger's work~\cite{Schrodinger1930}, where he proposed an improved relation
\begin{equation}\label{eq:SUR}
V_\rho(A)\, V_\rho(B) \ge \frac{1}{4} \left| \langle [A, B] \rangle_\rho \right|^2 + {\rm Cov}_\rho(A, B)^2, 
\end{equation}
by supplementing the lower bound with the term involving the symmetric quantum covariance ${\rm Cov}_\rho(A, B) \defeq \langle \{ A,  B \} \rangle_\rho/2 - \langle A \rangle_\rho \langle B \rangle_\rho$ dictated by the anticommutator \(\{A, B\} \coloneqq AB + BA\) of the observables concerned. 
\GEN{Thus, Schr\"odinger's refinement reveals that uncertainty trade-offs are shaped not only by the noncommutativity of observables, but also by their statistical correlation in the given state.
Unlike the commutator term, this covariance contribution is statistical in nature and reflects a kind of uncertainty trade-off that already exists even in classical probability theory.

Other recent developments include the improvement to the Kennard--Robertson relation for pure states proposed by Maccone and Pati~\cite{Maccone}}, which triggered subsequent developments~\cite{Wang,SONG20162925,Fan2020} (see also the earlier attempt by \cite{Yichen}). 
However, these formulations are primarily tailored to pure states and do not straightforwardly extend to mixed states without additional technical refinements~\cite{ChenFei2015SumUR,ChenCaoFeiLong2016VarianceUR,Fan2020}.  Meanwhile, Luo~\cite{Luo2} and Park~\cite{Park2005},  independently of each other, achieved improvements to the relation for mixed states by decomposing the variances into the classical and quantum contributions by drawing on insights from the Wigner--Yanase skew information~\cite{wigner1963}.  

Apart from these, other efforts sought modified relations involving sums of variances~\cite{Yichen,Maccone,Wang,SONG20162925,Fan2020} and the characterization of uncertainty regions~\cite{Li2015,Busch2019,Zhang2021,math4010008}.  In this regard, a systematic method of generating certain classes of both the product and the sum forms of trade-off relations has been investigated~\cite{Zheng_2020}.  The conditions for the equality of the uncertainty relation to hold are examined by Mal, Pramanik, and Majumdar~\cite{MPM}, as well as Zheng and Zhang~\cite{ZhengZhang2017} on the two-level system.  Specifically, these conditions have been accounted for in terms of the purity (or linear entropy) of the state, which offers a method to quantify the mixedness based on measurable values.  For the position and momentum observables, Hall derived an exact uncertainty relation~\cite{HallExact}.

Formulations involving alternative measures of uncertainty have also been explored, including entropic measures~\cite{Deutsch,Maassen,Kraus_1987,Berta_2010,RevModPhys.89.015002,HallRenyi}, Wigner--Yanase skew information~\cite{LZ2004,Luo2,Yanagi_2005,LZ2005,Kosaki_2005,Li_2009,FURUICHI2009179,Yanagi_2010,Chen_2016}, maximum probabilities~\cite{Landau1961,MiyaderaImai}, Fisher information~\cite{Gibilisco_2007,FSG2015,CG2022,TGFF2022}, and quantum coherence~\cite{SPB2016,Liu2016,Plenio2,PhysRevA.96.032313,Rastegin2017,Luo_2019}. 
For comprehensive reviews, the reader is referred to Refs.~\cite{Peres1995,Wehner_2010,busch2016quantum,HilgevoordUffink2024,Englert}.

\GEN{Among these refinements, a particularly simple and elegant extension of the Robertson--Schr\"odinger relation is the relation proposed by Hayashi (see Eq.~(7.27) and Exercise {\bf 7.14} in \cite{Hayashi2016QuantumInfo})\footnote{ 
The following equivalent form,
\[
\det\left(
\begin{array}{cc}
\operatorname{Cov}_{\rho}(A,A) & \operatorname{Cov}_{\rho}(A,B) \\
\operatorname{Cov}_{\rho}(A,B) & \operatorname{Cov}_{\rho}(B,B)
\end{array}
\right)
\geq
\left(
\frac{\operatorname{tr}\left|\sqrt{\rho}[A,B]\sqrt{\rho}\right|}{2}
\right)^2,
\]
is given as Exercise {\bf 7.14} in \cite{Hayashi2016QuantumInfo}. Historically, this relation was proved by Hayashi using a technique developed by Nagaoka \cite{Nagaoka2005CRbound}.
}:
\begin{equation}\label{eq:NH}
V_\rho(A)\, V_\rho(B) \ge \frac{1}{4}
\left(
\Tr\Bigl|\sqrt{\rho} [A, B] \sqrt{\rho}\Bigr|
\right)^2 + {\rm Cov}_\rho(A, B)^2, 
\end{equation}
where $|X| := \sqrt{X^\dagger X}$ denotes the absolute value, or modulus, of $X$.
This relation provides a tighter bound than the original Schr\"odinger relation for mixed states, while reducing to it when the state is pure. In particular, the relation becomes an exact equality relation for qubit systems (see Exercise {\bf  7.15} in \cite{Hayashi2016QuantumInfo}). 
However, the lower bound in Eq.~\eqref{eq:NH} involves the trace of $|\sqrt{\rho}[A,B]\sqrt{\rho}|$, and is not expressed simply as the expectation value of a fixed observable of the quantum system. (Notice here that the presence of the modulus prevents us from using the cyclicity of the trace to rewrite this term in a simple, operationally meaningful form, such as an expectation value $\langle X\rangle_\rho = \Tr(\rho X)$ for some fixed observable $X$.) 
This should be contrasted with an important feature of the Robertson and Schr\"odinger relations: their lower bounds are expressed through expectation values of observables, or quantities directly constructed from them.
This feature is physically significant, since it makes the bounds closely connected to experimental verification, in contrast to lower bounds formulated only in terms of more abstract mathematical quantities.
Thus, although Eq.~\eqref{eq:NH} is a mathematically powerful refinement, its direct operational interpretation as an uncertainty relation is somewhat less transparent than that of the Robertson and Schr\"odinger relations.}

\paragraph*{This work.}
\GEN{With these considerations in mind, we establish in this paper a universal extension of the Robertson--Schr\"odinger relation by adding a simple trade-off term to the Robertson--Schr\"odinger bound.
Crucially, this additional term is not merely an abstract mathematical lower bound, but is expressed as the expectation value of a positive observable, and therefore retains a direct connection to experimental verification.
More specifically, it explicitly involves the commutator and captures a hidden quantum contribution to uncertainty that is missed by the conventional Robertson--Schr\"odinger relation.
The significance of this improvement is manifested in various examples: even when a genuine trade-off exists, the standard Robertson--Schr\"odinger bound can fail to detect it, whereas our relation captures such hidden trade-offs particularly effectively for mixed states.
Finally, for two-level systems, our relation gives an exact equality for the product of variances, rather than merely providing a lower bound.}

\section*{Beyond Robertson and Schr\"odinger Relations \\
 {\fontsize{13pt}{12pt}\selectfont $\qquad$--- Revealing Hidden Quantum Uncertainty}}

\GEN{We now state our main uncertainty relation. 
For arbitrary observables $A$ and $B$ and an arbitrary quantum state $\rho$, the following uncertainty relation holds:
\begin{equation}\label{eq:UR_2New}
V_\rho(A) V_\rho(B) \geq
\underbrace{\frac{1}{4}\bigl|\Exp{[A,B]}_\rho\bigr|^2}_{\text{Robertson term}}
+
\underbrace{{\rm Cov}_\rho(A, B)^2}_{\text{Schr\"odinger term}}
+
\underbrace{\frac{\lm\lSm}{\lm+\lSm} \langle \bigl|[A,B]\bigr|^2\rangle_\rho}_{\text{New Trade-off}} ,
\end{equation}
where $\lambda_1 \leq \lambda_2$ are the smallest and second-smallest eigenvalues of the density matrix $\rho$ and $\bigl|[A,B]\bigr|^2 := [A,B]^\dagger [A,B]$ denotes the squared modulus of the commutator.
} This relation extends the Robertson--Schr\"odinger inequality~\eqref{eq:SUR} by incorporating an additional trade-off term involving the commutator $[A,B]$, thereby capturing a contribution to quantum uncertainty that arises from noncommutativity and is overlooked in the traditional formulation. 
Another feature of the new term is its dependence on the smallest and second-smallest eigenvalues of the density matrix, which strengthens the bound particularly for mixed states, reflecting how the degree of mixing amplifies the quantum uncertainty.
Although, at first glance, this specific dependence on individual eigenvalues---rather than more conventional quantities such as purity or entropy---may seem artificial or purely mathematical, \GEN{we show that it is in fact optimal: the coefficient in front of the expectation value of the modulus of the commutator is the largest possible, and hence the resulting inequality cannot be further improved within this form.}

Remarkably, for two-level quantum systems, i.e., qubit systems, the inequality becomes an \emph{exact equality} for any state and any pair of observables, rather than merely providing a lower bound.
In this case, the following exact uncertainty relation holds:
\begin{equation}\label{eq:UR_3}
V_\rho(A)\, V_\rho(B) = 
\underbrace{\frac{1}{4} \bigl| \langle [A,B] \rangle_\rho \bigr|^2}_{\text{Robertson Term}} + 
\underbrace{{\rm Cov}_\rho(A, B)^2}_{\text{Schr\"odinger Term}} + 
\underbrace{\GEN{\frac{1 - P(\rho)}{2} \langle \bigl|[A,B]\bigr|^2\rangle}}_{\text{New Trade-off}},
\end{equation}
\GEN{where \(P(\rho):=\Tr\rho^2\) is the purity of \(\rho\).
Here we intentionally omit the subscript $\rho$ from the expectation-value part of the last term, since, for two-level systems, this factor becomes state-independent and coincides with the squared Frobenius norm of the commutator: $ \langle \bigl|[A,B]\bigr|^2\rangle_\rho = \frac{1}{2}\|[A,B]\|^2:= \frac{1}{2}\Tr[A,B]^\dagger [A,B]$.
This separation between the state-dependent prefactor and the state-independent commutator term cleanly disentangles the effects of mixedness and noncommutativity, thereby removing an ambiguity in conventional uncertainty trade-offs, where state dependence can obscure, or even suppress, the contribution arising from noncommutativity.}

\section*{Methods}

\GEN{We proceed to prove the new uncertainty relation~\eqref{eq:UR_2New} and the exact relation~\eqref{eq:UR_3} for two-level systems.
The main idea of the proof is as follows.
First, we extend the B\"ottcher--Wenzel inequality to a state-dependent form.
This proves the uncertainty relation conjectured in Ref.~\cite{MKHC2}.
We then incorporate the Schr\"odinger covariance term into this bound, which leads to the new uncertainty relation~\eqref{eq:UR_2New}.
Finally, the exact relation~\eqref{eq:UR_3} for two-level systems is obtained by direct computation.
}

\GEN{Before giving the proof, we note that the new trade-off term in~\eqref{eq:UR_2New} vanishes when $\lambda_1=0$.
In this case, our relation reduces to the conventional Robertson--Schr\"odinger relation. 
The same observation applies to the relevant infinite-dimensional setting, where the bottom of the spectrum is zero: the additional term vanishes, and the asserted inequality again reduces to the conventional Robertson--Schr\"odinger relation. It is therefore sufficient to prove the relation \eqref{eq:UR_2New} for quantum systems with a $d$-dimensional Hilbert space $\CA^d$ and strictly positive density matrices. In what follows, let $\rho>0$ be such a faithful state, and let $A$ and $B$ be arbitrary observables represented by Hermitian matrices on $\CA^d$.
We arrange the eigenvalues of $\rho$ in ascending order:
\[
0<\lambda_1 \leq \lambda_2 \leq \cdots \leq \lambda_d.
\]}

\bigskip 

\paragraph*{A state-dependent B\"ottcher--Wenzel inequality and an associated uncertainty relation} 

Let us begin by recalling the celebrated B\"ottcher-Wenzel inequality \cite{Bottcher2005,VongJin2008BottcherWenzel,BOTTCHER20081864,Audenaert2010VarianceBounds} which provides the following tight bound for the commutator of arbitrary $d \times d$ complex matrices $X$ and $Y$

\begin{equation}  \label{BW}
    \| [X,Y]\|^2 \leq 2 \|X\|^2 \|Y\|^2 ,
\end{equation}
where $\|X\|= \sqrt{{\rm Tr} X^\dagger X}$ denotes the Frobenius, or Hilbert--Schmidt, norm.
To generalize (\ref{BW}) fix an arbitrary $\omega > 0$ and define the new $\omega$-inner product in $\Md$
\begin{equation}
   \braket{X}{Y}_\omega:= {\rm Tr}(\omega X^\dagger Y) ,
\end{equation}
and the corresponding $\omega$-norm via $\|X\|_\omega := \sqrt{ \braket{X}{X}_\omega }$. If $\omega = \mathbb{I}$, the identity matrix, it reduces to the original Frobenius (or Hilbert-Schmidt) norm $\|X\|= \sqrt{{\rm Tr} X^\dagger X}$. 

\GEN{A key ingredient in the proof of~\eqref{eq:UR_2New} is the following weighted commutator inequality, a generalization of the B\"ottcher--Wenzel inequality: for any complex matrix $X$ and for any normal matrix $Y$, 
\begin{equation}\label{BW-new}
    \| [X,Y]\|_\omega^2
    \leq
    \frac{\omega_1 + \omega_2}{\omega_1 \omega_2}
    \|X\|_\omega^2 \|Y\|_\omega^2 ,
\end{equation}
where $0 < \omega_1 \leq \omega_2 \leq \cdots \leq \omega_d$ are the eigenvalues of $\omega$.
This inequality was originally conjectured in Ref.~\cite{MKHC} for general matrices $X$ and $Y$. 
In this paper, we prove this in the case where one of the two matrices is normal, which in particular covers the Hermitian case considered here. The proof is given in the Supplementary Information.}
Clearly, if $\omega=\mathbb{I}$, then (\ref{BW-new}) reduces to (\ref{BW}). 

\GEN{By specializing $\omega$ to the density matrix $\rho$ and $X,Y$ to Hermitian matrices $A,B$ in~\eqref{BW-new}, we obtain the following simple uncertainty relation:
\begin{equation}\label{eq:FamUR}
  V_\rho(A) V_\rho(B)
  \geq
  \frac{\lm\lSm}{\lm+\lSm}\langle \left|[A, B]\right|^2 \rangle_\rho,
\end{equation}
where $\lm$ and $\lSm$ are the smallest and second-smallest eigenvalues of $\rho$, respectively. 
Indeed, let $\zExp{A} := A - \langle A \rangle_\rho \mathbb{I}$ denote the shifted operator with zero mean in the state $\rho$.
Then
\begin{equation}\label{eq:ShAV}
    \|\zExp{A}\|^2_\rho
    =
    \Tr\left((A - \langle A \rangle_\rho \I)^2 \rho\right)
    =
    V_\rho(A).
\end{equation}
Since $\zExp{A}$ and $\zExp{B}$ are Hermitian and hence normal, and since $[\zExp{A}, \zExp{B}] = [A, B]$, applying~\eqref{BW-new} with $X=\zExp{A}$, $Y=\zExp{B}$, and $\omega=\rho$ yields the following relation:
\begin{equation}\label{eq:URconjPrev}
  V_\rho(A) V_\rho(B)
  \geq
  \frac{\lm\lSm}{\lm+\lSm} \|[A, B]\|^2_\rho. 
\end{equation}
Finally, noting that
\begin{equation}
\| [A,B]\|^2_\rho
=
\Tr [A,B]^\dagger [A,B] \rho
=
\Tr \bigl|[A,B]\bigr|^2 \rho
=
\langle \bigl|[A,B]\bigr|^2 \rangle_\rho,\label{eq:NormExp}
\end{equation}
we obtain the relation~\eqref{eq:FamUR}.

The relation \eqref{eq:URconjPrev} was conjectured in Ref.~\cite{MKHC} on the basis of numerical evidence. Its physical significance and comparison with other uncertainty relations were investigated in detail in Ref.~\cite{MKHC2}. 
In the present work, we have proved this rigorously as a consequence of the state-dependent B\"ottcher--Wenzel inequality~\eqref{BW-new}. 
Moreover, by rewriting it in the equivalent form~\eqref{eq:FamUR}, we have recast the relation in a physically more transparent form.}

\GRN{As already pointed out in Ref. [66], the relation~\eqref{eq:FamUR} is not merely one possible refinement of the Robertson--Schr\"odinger relation, but is optimal within the natural class of uncertainty relations of the form
\[
  V_\rho(A) V_\rho(B)
  \geq
  b(\rho) \left\langle\left|[A, B]\right|^2 \right\rangle_\rho,
\]
where $b(\rho)$ is a positive constant depending only on $\rho$.
Since $b(\rho)$ appears as the coefficient of the lower bound, a larger value of $b(\rho)$ gives a stronger inequality.
Thus, the optimal coefficient is the largest value of $b(\rho)$ for which the above inequality holds universally for all observables $A$ and $B$.

More precisely, it turns out that, among all constants $b(\rho)$ for which the above inequality holds universally for all observables $A$ and $B$, the largest possible value is
\[
b_{\rm opt}(\rho):= \frac{\lm\lSm}{\lm+\lSm}.
\]
Thus, the coefficient appearing in~\eqref{eq:FamUR} cannot be improved in general. 
To see this, it is enough to exhibit observables $A$ and $B$ that attain equality in~\eqref{eq:FamUR} with a nonzero variance product.
Such a simple example is given by
\begin{equation}
A:=\lambda_2\ketbra{1}{1}-\lambda_1\ketbra{2}{2},
\qquad
B:=\ketbra{1}{2}+\ketbra{2}{1}, \label{eq:Optimal}
\end{equation}
where $|1\rangle$ and $|2\rangle$ are orthonormal eigenvectors of $\rho$ corresponding to its smallest and second-smallest eigenvalues $\lambda_1$ and $\lambda_2$, respectively. 
Indeed, a direct computation shows
\[
V_\rho(A) =\lambda_1\lambda_2(\lambda_1+\lambda_2), V_\rho(B)
=
\lambda_1+\lambda_2,
\]
and 
\[
\left\langle |[A,B]|^2 \right\rangle_\rho
=
\Tr\left(\rho |[A,B]|^2\right)
=
(\lambda_1+\lambda_2)^3.
\]
\RED{Consequently, the equality in~\eqref{eq:FamUR} is attained}. 
This suggests that the dependence on the smallest and second-smallest eigenvalues of $\rho$ in the uncertainty relation~\eqref{eq:FamUR} --- and, in particular, the absence of any dependence on the remaining eigenvalues --- is not merely a mathematical artifact, but has a structural origin in quantum theory.
}

\begin{Remark}
Let us observe the elementary comparison
\begin{equation}\label{relRFnorms}
    \sqrt{\lambda_1} \|X\| \leq \|X\|_\rho \leq \sqrt{\lambda_d} \|X\|, 
\end{equation}
which follows directly from $\lambda_1 \I \leq \rho \leq \lambda_d \I$. 
Together with the B\"ottcher--Wenzel inequality, this yields
\[
\|[X,Y]\|_\rho^2
\leq
\lambda_d \|[X,Y]\|^2
\leq
2\lambda_d \|X\|^2 \|Y\|^2
\leq
\frac{2\lambda_d}{\lambda_1^2}
\|X\|_\rho^2 \|Y\|_\rho^2 .
\]
Therefore, by the same argument as above, we obtain the following uncertainty relation
\begin{equation}\label{eq:LooseUR}
 V_\rho(A) V_\rho(B)
 \geq
 \frac{\lambda_1^2}{2\lambda_d}\langle\left|[A, B]\right|^2 \rangle_\rho.
\end{equation}
This relation was shown in Refs.~\cite{MKHC,MKHC2}, but the bound is not particularly tight.
In contrast, the relation~\eqref{eq:FamUR} provides a sharp bound.
\end{Remark}

\paragraph*{Uncertainty Relation (\ref{eq:UR_2New}).} 

\GEN{We are now ready to prove the uncertainty relation~\eqref{eq:UR_2New}, which may be viewed as a strengthened form of the Robertson--Schr\"odinger relation incorporating the additional commutator contribution in~\eqref{eq:FamUR}. 
The uncertainty relation~\eqref{eq:UR_2New} also follows from the generalized inequality \eqref{BW-new} together with the following simple observation. 
Suppose that the inequality
\[
c(\omega)\|X\|_\omega^2\|Y\|_\omega^2 \geq \|[X,Y]\|_\omega^2
\]
holds for every complex matrix $X$ and every normal matrix $Y$.
Then the stronger inequality
\begin{equation}\label{eq:strongUR}
c(\omega) \left( \|X\|_\omega^2 \|Y\|_\omega^2 - |\braket{X}{Y}_\omega|^2 \right)
\geq \|[X, Y]\|_\omega^2
\end{equation}
also holds. 
The proof is given in the Supplementary Information.
Therefore, applying this observation to~\eqref{BW-new}, we obtain the following inequality, especially for any Hermitian matrices $X$ and $Y$:
\begin{equation}\label{eq:XYst1}
\|X\|_\omega^2 \|Y\|_\omega^2
\geq |\Tr(X Y \omega)|^2 + 
\frac{\omega_1 \omega_2}{\omega_1 + \omega_2}
\|[X, Y]\|_\omega^2.
\end{equation}
Using the relation
\begin{eqnarray*}
|\Tr (XY \omega)|^2 =  \frac{1}{4} \bigl(\bigl|\Tr \{X,Y\} \omega\bigr|^2 + \bigl|\Tr [X,Y] \omega \bigr|^2\bigr),
\end{eqnarray*}
which follows from the decomposition $XY = \frac{\{X,Y\}}{2} + i \frac{[X,Y]}{2i}$, one obtains 
\begin{equation}\label{eq:XYst2}
\|X\|_\omega^2 \|Y\|_\omega^2
\geq \frac{1}{4} \bigl|\Tr [X,Y] \omega \bigr|^2 + \bigl(\frac{1}{2}\Tr \{X,Y\} \omega\bigr)^2   + 
\frac{\omega_1 \omega_2}{\omega_1 + \omega_2}
\|[X, Y]\|_\omega^2.
\end{equation}
Now, specializing this inequality to the case where $\omega$ is the density operator $\rho$, and setting
$X=\zExp{A}$ and $Y=\zExp{B}$ for observables $A$ and $B$, and noting that $[\zExp{A},\zExp{B}] = [A,B]$, $\frac{1}{2} \langle\{\zExp{A},\zExp{B}\}\rangle_\rho = {\rm Cov}_\rho(A,B)$ and \eqref{eq:ShAV}, one obtains the following uncertainty relation
\begin{equation}\label{eq:strongUR2zver}
V_\rho(A) V_\rho(B) 
\geq \frac{1}{4} \bigl| \langle [A,B] \rangle_\rho \bigr|^2 + {\rm Cov}_\rho(A,B)^2 + \frac{\lambda_1 \lambda_2}{\lambda_1 + \lambda_2}
\|[A, B]\|_\rho^2.
\end{equation}
Together with~\eqref{eq:NormExp}, this completes the derivation of the uncertainty relation~\eqref{eq:UR_2New}.

\GRN{
The same example in \eqref{eq:Optimal} also shows that the stronger relation~\eqref{eq:UR_2New} is tight in the relevant sense.
Indeed, for the above choice of $A$ and $B$, one has
\[
\langle [A,B]\rangle_\rho=0,
\qquad
{\rm Cov}_\rho(A,B)=0,
\]
while the additional term satisfies
\[
\frac{\lambda_1\lambda_2}{\lambda_1+\lambda_2}
\left\langle |[A,B]|^2\right\rangle_\rho
=
V_\rho(A)V_\rho(B).
\]
Thus equality is also attained in~\eqref{eq:UR_2New}.
In particular, the coefficient of the last term in~\eqref{eq:UR_2New} cannot be increased in general, even after the Robertson and covariance terms are included.

This reinforces the point that the dependence on the smallest and second-smallest eigenvalues of $\rho$ is not an accidental feature of the proof.
Rather, these two eigenvalues determine the optimal strength of the additional noncommutative contribution in this class of state-dependent uncertainty relations.
In this sense, the absence of the remaining eigenvalues in the coefficient reflects a genuine structural aspect of the quantum uncertainty trade-off, rather than a merely technical artifact.
}

\begin{Remark} Note, that due to $\lambda_1 \leq \lambda_2$ one has
\[   \frac{\lambda_1}{2} \leq \frac{\lambda_1 \lambda_2}{\lambda_1 + \lambda_2} \leq \frac{\lambda_2}{2} ,\]
and hence the uncertainty relation~\eqref{eq:UR_2New} implies 

\begin{equation}\label{lambda/2}
V_\rho(A) V_\rho(B) \ge\ \frac{1}{4}|\Exp{[A,B]}_\rho|^2 + {\rm Cov}_\rho(A, B)^2 + \frac{\lambda_1}{2} \langle \bigl|[A,B]\bigr|^2\rangle_\rho .
\end{equation}
Clearly, it is weaker than \eqref{eq:UR_2New},
however, it requires the knowledge of the smallest eigenvalue only. Of course if $\lambda_2=\lambda_1$ the original relation \eqref{eq:UR_2New} reduces to (\ref{lambda/2}).
\end{Remark}
}

\paragraph*{Uncertainty Relation \eqref{eq:UR_3} for qubit systems.} As mentioned above, the uncertainty relation \eqref{eq:UR_2New} turns out to yield an exact equality~\eqref{eq:UR_3} for the product of variances in the case of two-level systems. 

To see this, it is convenient to use the Bloch vector representation for the state:
\begin{equation}\label{Bloch}
\rho = \frac{1}{2}(\mathbb{I} + \bm{r} \cdot \bm{\sigma}) := \frac{1}{2}(\mathbb{I} + \sum_{i=1}^3 r_i \sigma_i),
\end{equation}
where $\bm{r} = (r_1, r_2, r_3) \in \mathbb{R}^3$ with $|\bm{r}| \le 1$, and $\bm{\sigma} = (\sigma_1, \sigma_2, \sigma_3) = (\sigma_x, \sigma_y, \sigma_z)$ denotes the vector of Pauli matrices. We also expand the observables $A$ and $B$ as
\begin{equation}\label{AB}
A=a_0\I+\bm a\cdot\bm \sigma, \quad B=b_0\I+\bm b\cdot\bm \sigma
\end{equation}
with $a_0, b_0 \in \mathbb{R}$, $\bm a=(a_1,a_2,a_3), \bm b=(b_1,b_2,b_3) \in \mathbb{R}^{3}$. 
Here, and in what follows, we use the following standard notations.  
For three-dimensional vectors ${\bm a} = (a_1, a_2, a_3)$ and ${\bm b} = (b_1, b_2, b_3)$ in $\mathbb{R}^3$,  
we denote the Euclidean inner product, Euclidean norm, and cross product by
$\bm{a} \cdot \bm{b} := \sum_{i=1}^3 a_i b_i$, $|\bm{a}| := \sqrt{\sum_{i=1}^3 a_i^2}$, and $\bm{a} \times \bm{b} := (a_2 b_3 - a_3 b_2,\, a_3 b_1 - a_1 b_3,\, a_1 b_2 - a_2 b_1)$, respectively. 
The details of all calculations appearing below are provided in the Supplementary Information.

In two-level systems, it holds that $\lambda_1 + \lambda_2 = 1$ and $\lambda_1 \lambda_2 = \frac{(\lambda_1+\lambda_2)^2- (\lambda_1^2+\lambda_2^2)}{2} = \frac{1-P(\rho)}{2}$.
Interestingly, in this case, the $\rho$-norm of the commutator becomes independent of the state and essentially coincides with the Frobenius norm:
\GEN{A direct computation shows 
\begin{equation}\label{iesNFN}
\langle \bigl| [A,B]\bigr|^2 \rangle_\rho = \Tr [A,B]^\dagger [A,B] \rho = 4 |\bm{a}\times\bm{b}|^2
\end{equation}
which is independent of the state. 
In particular, this coincides with the value for the maximally mixed state $\rho_{\max}=\I/2$, and hence  
\[
\langle \bigl| [A,B]\bigr|^2\rangle_\rho = \frac{1}{2}\|[A,B]\|^2.
\]
As mentioned above, this was the reason why we omitted the explicit $\rho$-dependence from the additional expectation-value term in~\eqref{eq:UR_3} and wrote it as
\begin{equation}\label{Bdd2lev}
  \frac{\lm\lSm}{\lm+\lSm} \langle \bigl|[A,B]\bigr|^2\rangle_\rho = \frac{1 - P(\rho)}{2} \langle \bigl| [A,B]\bigr|^2 \rangle.
\end{equation}}
Now, straightforward computations yield
\begin{equation}\label{VAVB}
V_\rho(A)= |\bm a|^2-(\bm a\cdot\bm{r})^2, V_\rho(B) = |\bm b|^2-(\bm b\cdot\bm{r})^2
\end{equation}
and 
\begin{subequations}\label{bounds}
\begin{align}
\frac{1}{4}|\langle[A,B]\rangle_\rho|^2 &= |(\bm a\times\bm b)\cdot\bm{r}|^2, \\
{\rm Cov}_\rho(A,B)^2 &= |\bm a\cdot\bm b-(\bm a\cdot\bm{r})(\bm b\cdot\bm{r})|^2, \\
\frac{1 - P(\rho)}{2} \langle \bigl| [A,B]\bigr|^2 \rangle &= (1-|\bm{r}|^2)|\bm a\times \bm b|^2. \label{eq:thirdeq}
\end{align}
\end{subequations}
Using these expressions, together with the elementary identities $|(\bm a\times\bm b)\cdot\bm{r}|^2 + |(\bm a\times\bm b)\times \bm{r}|^2 = |(\bm a\times\bm b)|^2 |\bm{r}|^2$, $|\bm a\times \bm b|^2 = |\bm a|^2 |\bm b|^2 - ({\bm a} \cdot {\bm b})^2$
and the vector triple product identity $(\bm{a} \times \bm{b}) \times \bm{r} = (\bm{a} \cdot \bm{r}) \bm{b} - (\bm{b} \cdot \bm{r}) \bm{a}$,
one verifies, by a direct computation, the equality relation~\eqref{eq:UR_3}.

\begin{Remark}
\GEN{
Exact equality relations for two-level systems have already been studied in several works. 
As mentioned above, Hayashi's relation~\eqref{eq:NH} becomes an exact equality for two-level systems. 
In Ref.~\cite{MPM}, the difference between the left- and right-hand sides of the Schr\"odinger relation~\eqref{eq:SUR} for two-level systems was shown to be
\begin{equation}\label{MPM}
\bigl(|{\bm a}|^2 |{\bm b}|^2 - ({\bm a} \cdot {\bm b})^2\bigr) S_l(\rho),
\end{equation}
where $S_l(\rho) := 2(1-P(\rho))$ denotes the linear entropy of the state $\rho$.
This difference corresponds to our additional trade-off term~\eqref{Bdd2lev}. 
Independently, Ref.~\cite{ZhengZhang2017} characterized the same contribution by
\begin{equation}\label{Zheng}
\frac{1-P(\rho)}{8}(\xi(A,A)\xi(B,B)-\xi(A,B)^2),
\end{equation}
where $\xi(X,Y) := 2\Tr (XY) -\Tr(X) \Tr(Y) = 4{\rm Cov}_{\rho_{\max}}(X,Y)$. 
Indeed, it is straightforward to see that these can be recovered from our equality~\eqref{eq:UR_3}, and are thereby understood as specific instances of our new uncertainty relation~\eqref{eq:UR_2New}: It is easy to see that~\eqref{eq:thirdeq} coincides with~\eqref{MPM}. 
On the other hand, showing that~\eqref{eq:thirdeq} also coincides with~\eqref{Zheng} requires an elementary but somewhat tedious calculation, which is provided in the Supplementary Information. A noteworthy distinction, however, is that while their bounds also highlight a trade-off characterized by the purity of the state, our formulation explicitly demonstrates that this term arises from the commutator of observables, thereby identifying the trade-off as being of quantum origin.  
An even more striking distinction is that the uncertainty relation \eqref{eq:UR_2New} holds not only for two-level systems, but also for arbitrary quantum systems.  
In the general case, the contribution of mixedness to the uncertainty relation is not determined simply by the purity itself, but rather by the smallest and the second-smallest eigenvalues of the density matrix.}
\end{Remark}


\begin{Remark}
    \JAEHA{
Although our relation \eqref{eq:UR_2New} becomes an exact equality for two-level systems, it is still instructive to look at the relation~\eqref{eq:FamUR} separately. 
For two-level systems, it takes the form
\begin{equation}
    \left( |{\bm a}|^{2} - ({\bm a}\cdot{\bm r})^{2} \right)
    \left( |{\bm b}|^{2} - ({\bm b}\cdot{\bm r})^{2} \right)
    \geq
    (1 - |{\bm r}|^{2}) |{\bm a} \times {\bm b}|^{2},
\end{equation}
which makes the equality conditions particularly transparent.  We first note that whenever either observable is degenerate (\textit{i.e.}, ${\bm a} = 0$ or ${\bm b} = 0$), the relation~\eqref{eq:FamUR} reduces to the uninteresting inequality $0 \geq 0$.  Apart from this trivial case, one immediately obtains the following conditions: \GRN{For nondegenerate (\textit{i.e.}, $\bm{a}, \bm{b} \neq 0$) two-level observables, we find}
\begin{itemize}
    \item For the maximally mixed state, $|{\bm r}|=0$, equality holds if and only if the observables are orthogonal, namely
    \[
    {\bm a}\perp{\bm b} \iff {\bm a}\cdot{\bm b}=0,
    \]
    \GRN{that is, the corresponding nondegenerate two-level observables are complementary, or equivalently, their eigenbases are mutually unbiased.}

    \item For a pure state, $|{\bm r}|=1$, equality holds if and only if the state $\rho$ is an eigenstate of either $A$ or $B$, or equivalently,
    \[
    \frac{{\bm a}}{|{\bm a}|} = \pm {\bm r}
    \qquad \text{or} \qquad
    \frac{{\bm b}}{|{\bm b}|} = \pm {\bm r}.
    \]

    \item \GRN{For mutually unbiased observables, namely,} ${\bm a}\perp{\bm b}$, equality holds if and only if the state commutes with either observable, or equivalently
    \[
    {\bm a}\parallel{\bm r}
    \qquad \text{or} \qquad
    {\bm b}\parallel{\bm r}.
    \]

    \item For commuting observables, ${\bm a}\parallel{\bm b}$, equality holds if and only if the state $\rho$ is an eigenstate of either, hence both, $A$ and $B$, or equivalently
    \[
    \frac{{\bm a}}{|{\bm a}|} = \pm {\bm r}
    \qquad \text{and} \qquad
    \frac{{\bm b}}{|{\bm b}|} = \pm {\bm r}.
    \]
\end{itemize}
}
\end{Remark}
\section*{Comparison and Outlook}
    
In this work, we have proven the previously conjectured uncertainty relation~\eqref{eq:FamUR}, and further shown that it leads to a stronger uncertainty relation~\eqref{eq:UR_2New}, which generalizes the Robertson--Schr\"odinger relation.  
What distinguishes our generalization from many others is that it incorporates a genuinely new trade-off term which explicitly involves the commutator of observables and, \GEN{at the same time, is expressed as the expectation value of a physical observable.}
This allows us to uncover a fundamental uncertainty trade-off of quantum origin---specifically, one rooted in non-commutativity---that is not captured by the conventional Robertson--Schr\"odinger relation.
Moreover, this relation becomes an exact equality~\eqref{eq:UR_3} for all two-level systems. 
To illustrate these improvements in a quantitative and concrete way, we first compare our bounds with the Robertson and Schr\"odinger bounds in two-level systems, and then demonstrate their strength by considering spin observables.
We then discuss possible experimental tests of our relations, and conclude the paper with some general remarks.

\begin{figure}[htbp]
\centering
\includegraphics[width=0.825\textwidth]{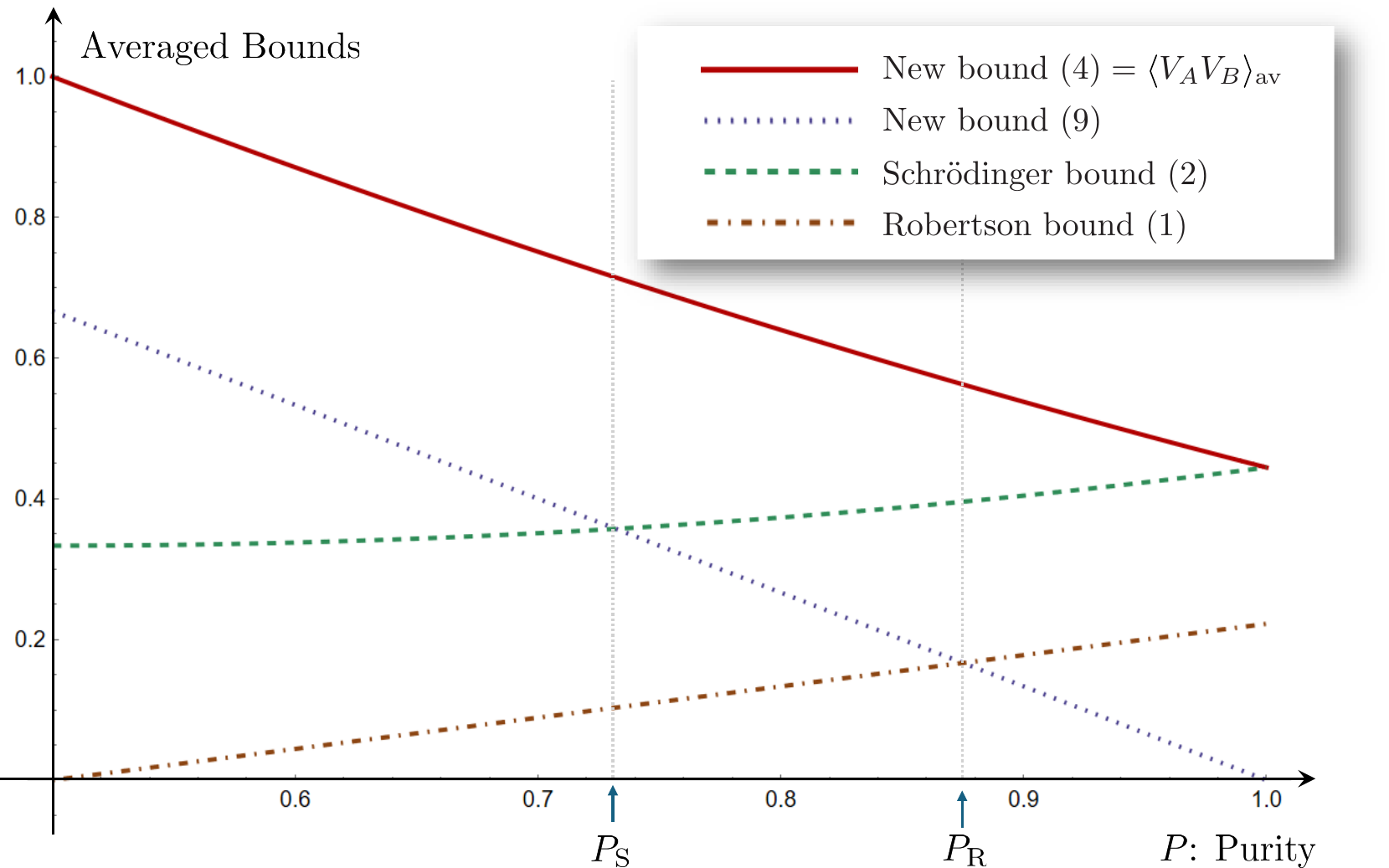}
\caption{The averaged bounds given by the Robertson (Brown, DotDashed), Schr\"odinger (Green, Dashed), and our uncertainty relations~\eqref{eq:FamUR} (Blue, Dotted) and~\eqref{eq:UR_2New} (Red, Solid), plotted as a function of the purity.  
Note that, in the qubit case, relation~\eqref{eq:UR_2New} becomes an equality~\eqref{eq:UR_3}, and thus coincides with the product of variances.
}\label{comparison}
\end{figure}

\paragraph*{Comparison.} In Ref.~\cite{MKHC}, focusing on two-level systems, the averages of each bound, taken uniformly over the spin directions of $A$ and $B$ (with respect to the Haar measure on $S^2$), have been computed as follows:
\begin{subequations}\label{avBdd}
\begin{align}
\Bigl\langle \frac{1}{4}|\langle[A,B]\rangle_\rho|^2 \Bigr\rangle_{\rm av} &= \frac{2}{9}(2P-1)\label{rob2}\\
\Bigl\langle {\rm Cov}_\rho(A,B)^2\Bigr\rangle_{\rm av} &= \frac{2}{9}(2P^2-4P+3), \label{sch2}\\
\Bigl\langle \frac{1 - P(\rho)}{2} \langle \bigl| [A,B]\bigr|^2 \rangle\Bigr\rangle_{\rm av}&=\frac{4}{3}(1-P), \label{ours2-2}
\end{align}
\end{subequations}
where $P= P(\rho)$. Using these expressions and the equality relation \eqref{eq:UR_3}, we also get  
\begin{equation}
\Bigl\langle V_\rho(A)V_\rho(B) \Bigr\rangle_{\rm av} = \frac{4}{9}(2-P)^2, \label{var}
\end{equation}
In Fig.~\ref{comparison}, the bounds by Robertson, Schr\"odinger, and our bounds \eqref{eq:FamUR} and \eqref{eq:UR_2New} are plotted as a function of the purity. As the degree of mixedness of the state increases, bound \eqref{eq:FamUR} alone already captures a tighter trade-off than the Robertson and Schr\"odinger bounds when $P \le P_R := 7/8 = 0.875$ and $P \le P_S := \sqrt{3}-1 \simeq 0.732$. The refined bound \eqref{eq:UR_2New} (equivalently, \eqref{eq:UR_3}), however, coincides with the product of variances, establishing the ultimate limit in this case. 

\paragraph*{Illustration with spin operators.} To illustrate the power of the new relation \eqref{eq:UR_2New}, consider spin operators~\cite{SakuraiNap} $\{J_1,J_2,J_3\}$ in $\CA^d$ satisfying the standard commutation relation 
        $$   [J_k,J_\ell] = i\hbar\, \epsilon_{k\ell m} J_m,
        $$
where $\epsilon_{k\ell m}$ denotes the Levi-Civita symbol. 
In the canonical basis $|j,m\rangle$ one has

\begin{equation}
    J_3|j,m\rangle = \hbar m |j,m\rangle \ , \ \ \ \ J_1 = \frac 12 (J_+ + J_-) \ , \ \  J_2 = \frac{1}{2i} (J_+ - J_-) \ ,
\end{equation}
where

\begin{eqnarray}
J_\pm |j,m\rangle =  \hbar \sqrt{(j \mp m)(j\pm m+1)} \,  |j, m \pm 1 \rangle , 
\end{eqnarray}
and $m \in \{-j,-j+1,\ldots,j-1,j\}$ together with $d = 2 j+1$. In the maximally mixed state $\rho_{\max} = \I/d$ one finds
\begin{equation}
    \langle [J_1,J_2] \rangle_{\rho_{\max}} = i\hbar \langle J_3 \rangle_{\rho_{\max}} = 0 ,  \ \  \langle J_1 \rangle_{\rho_{\max}} = 0 \ , \ \  \langle J_2 \rangle_{\rho_{\max}} = 0 ,
\end{equation}
and
\begin{equation}
     \langle \{J_1,J_2\} \rangle_{\rho_{\max}} = \frac{1}{2i}  \langle J_+J_+ + J_- J_- \rangle_{\rho_{\max}} = 0 . 
\end{equation}
It is therefore clear that the Robertson--Schr\"odinger relation trivializes to $V_{\rho_{\max}}(J_1) V_{\rho_{\max}}(J_2) \geq 0$, whereas the last term in (\ref{eq:UR_2New}) provides a nontrivial contribution
\begin{equation}
    V_{\rho_{\max}}(J_1) V_{\rho_{\max}}(J_2) \geq  \frac{\hbar^2}{2d^2}  {\rm Tr}\,J_3^2 = \frac{\hbar^4}{2d^2} \alpha_j, 
\end{equation}
where the constant $\alpha_j$ reads
\begin{equation}
    \alpha_j = \sum_{m=-j}^j m^2 = \frac{j(j+1)(2j+1)}{3} .
\end{equation}
In particular for a qubit system one has $j=\frac 12$ and hence

\begin{equation}
    V_{\rho_{\max}}(J_1) V_{\rho_{\max}}(J_2) \geq  \left(\frac{\hbar}{2}\right)^4 . 
\end{equation}

\paragraph*{Possible experimental tests.}
We remark that our new uncertainty relation offers a prospect for experimental verification, as it is formulated solely in terms of measurable expectation values associated with suitably chosen observables. 
Tight, state-independent preparation uncertainty relations for qubit observables, as derived by Abbott {\it et al.}~\cite{math4010008}, have already been experimentally confirmed using neutron polarimetry for both pure and mixed spin states with varying polarization~\cite{Hasegawa}. 
Their approach can be directly adapted to enable an experimental verification of our new uncertainty relation in two-level (qubit) systems~\cite{HasegawaPrivate}.
For higher-dimensional systems, such as three-level systems, a photonic qutrit appears to be a promising candidate for experimental realization. 
For experimental tests of preparation-type uncertainty relations, see for example Ref.~\cite{Xiao:17}, where a qutrit was realized by encoding three orthogonal modes of single photons: the horizontal polarization in the lower spatial mode, the horizontal polarization in the upper spatial mode, and the vertical polarization in the upper spatial mode.  
This setup may provide a feasible platform for testing our new uncertainty relation beyond qubit systems. 
More generally, any qubit platform that enables high-fidelity measurements of mixed states and their associated variances—including in multi-qubit registers—such as trapped ions~\cite{Wineland1998,Schmidt-Kaler_2003,Schindler_2013,PhysRevA.79.023419,PhysRevLett.111.180501,PhysRevA.104.L060402,An2022,Ma2023}, state-of-the-art superconducting circuits~\cite{Nakamura1999,PhysRevLett.112.190504,PhysRevApplied.7.054020,PhysRevApplied.10.034040,Opremcak2021}, NV-center electron spins~\cite{doi:10.1126/science.276.5321.2012,Jelezko2006,doi:10.1126/science.1139831,RevModPhys.92.015004}, photonic qubits~\cite{OBrien2009,Flamini2018}, or nuclear spin registers~\cite{Pla2013,RevModPhys.95.025003,Reiner2024}—would be equally well suited for the experimental verification of our uncertainty relations.

\paragraph*{Physical interpretation and general remarks.}
For general quantum systems, the new trade-off term vanishes when the state has a zero eigenvalue---particularly in the case of pure states.  
Therefore, the relevance of our trade-off becomes prominent in the case of mixed states, and more precisely, in faithful mixed states where $\lambda_1$ (and thus $\lambda_2$) are nonzero.  
Next, in infinite-dimensional systems, $\lambda_1$ tends to zero asymptotically, and our term again vanishes, reducing the relation to the standard Schr\"odinger relation. 
This may partly explain why such a genuine trade-off intrinsic to quantum theory had not been identified in earlier formulations, even as we mark a century since the advent of Heisenberg's matrix mechanics.
However, recent developments in quantum information science (see, e.g., Refs.~\cite{NC,Barnett2009,Hayashi2016QuantumInfo,Wilde2017}) have underscored the significance of finite-level systems and the treatment of mixed states.  
In this context, our uncertainty relation \GEN{\eqref{eq:UR_2New}, a simple yet powerful refinement of the Robertson--Schr\"odinger relation that is universal and experimentally accessible,} may be expected to offer a tighter and more informative trade-off for applications such as quantum cryptography and quantum computation. 

\GRN{It is worth emphasizing that the fact that our relation reduces to the conventional Robertson--Schr\"odinger relation in the pure-state case should not be regarded as a drawback of the result.
Rather, it identifies the regime in which the new contribution becomes physically relevant.
Indeed, our work is not aimed at classifying minimum-uncertainty states obtained by optimizing over all states.
Instead, it addresses a different question: for a given state $\rho$, especially a mixed state, and a given pair of observables $A$ and $B$, what physically meaningful contributions constitute the uncertainty product $V_{\rho}(A)V_{\rho}(B)$? For this question, the Robertson--Schr\"odinger relation may become trivial even when the observables do not commute.
By contrast, our additional term detects precisely this missing contribution.
It reveals, especially for mixed states, a genuinely noncommutative contribution to the uncertainty trade-off that is invisible to the usual Robertson--Schr\"odinger lower bound.
Moreover, it should be emphasized that this is not merely one possible additional lower bound.
Within the natural class of state-dependent bounds considered in this work, the coefficient of this additional commutator term is optimal and cannot be improved in general.
Thus, our relation provides not only a stronger lower bound, but also a sharper structural constraint on quantum-mechanical statistics.
}

\GRN{Finally, let us comment on the physical role of uncertainty relations of the Robertson--Schr\"odinger type.
They are not merely practical tools for computing variances: if the state and observables are known, the product $V_\rho(A)V_\rho(B)$ can be evaluated directly.
Rather, their significance lies in expressing universal structural constraints on the statistical behavior of quantum observables.
The Robertson relation connects variances with noncommutativity through the expectation value of the commutator, and the Schr\"odinger refinement further incorporates the symmetric covariance.
Our relation reveals an additional positive contribution,
\[
    \frac{\lambda_1\lambda_2}{\lambda_1+\lambda_2}
    \langle |[A,B]|^2\rangle_{\rho},
\]
which is not visible in the conventional Robertson--Schr\"odinger relation.
More broadly, it may also serve as a consistency test of the quantum-mechanical description itself: within the assumptions of the model, experimentally observed statistics violating this relation would not be compatible with quantum mechanics.
}

\paragraph*{Concluding remarks.}

\GRN{
In this paper, we have established a universal refinement of the Robertson--Schr\"odinger uncertainty relation by identifying an additional positive contribution to the uncertainty product.
This contribution is expressed as the expectation value of the positive observable $|[A,B]|^2$, and is therefore directly tied to the noncommutativity of the observables.
We have shown that the coefficient of this additional term is optimal within the natural class of state-dependent commutator-norm bounds, and that the resulting relation becomes an exact equality for all two-level systems.

Although the present work focuses on trade-off relations for the product of variances, it would be an interesting future problem to investigate whether the same idea of optimizing mixed-state contributions can be extended to other forms of uncertainty relations, such as entropic uncertainty relations or sum-of-variances relations.
In particular, it would be valuable to understand whether analogous state-dependent terms can be found that couple the mixedness of the state to the complementarity or incompatibility of the observables in those settings.

We hope that our findings will not only deepen the understanding of fundamental uncertainty trade-offs in quantum systems, but also stimulate further theoretical and experimental investigations into the structure of quantum uncertainty.
}

\section*{Acknowledgements}

G.K.\ and A.M.\ would like to thank Michael Hall, Michele Dall’Arno, Francesco Buscemi, and Izumi Tsutsui for the useful discussions and comments and Yuji Hasegawa for valuable discussions and insightful comments regarding the possible experimental realization. 
G.K., H.O., and J.L. were supported by JSPS KAKENHI
Grants Nos. 24K06873, 23K03147, and 22K13970, respectively. D.C. was supported by the Polish National Science Center under Project No. 2024/55/B/ST2/01781.

\section*{Author Contributions}

Gen Kimura conceived the project and, together with Aina Mayumi, derived the main results, particularly the uncertainty relations~\eqref{eq:FamUR}, \eqref{eq:UR_2New} and \eqref{eq:UR_3}, following discussions with Jaeha Lee and Dariusz Chru\'sci\'nski. 
Hiromichi Ohno played a leading role in the proof of Theorem \ref{thm1}.
Dariusz Chru\'sci\'nski provided the illustration with spin operators. 
All authors participated in writing and reviewing the final version.

\bigskip 

\appendix 
\renewcommand{\thesection}{\Alph{section}} 
\counterwithin{equation}{section}   

\section*{Supplementary Information}

\section{Key Theorem for the Proof of the Uncertainty Relation \eqref{eq:FamUR}}

\GEN{In this section, we prove the inequality~\eqref{BW-new}.
Although the application to~\eqref{eq:FamUR} only requires the case where both $X$ and $Y$ are Hermitian, we establish the inequality in a slightly more general setting: one of the two matrices is allowed to be arbitrary complex, while the other is assumed to be normal. 
As will be shown in the next section, this generalization is essential for deriving the inequality~\eqref{eq:strongUR}, which in turn is needed to prove the general uncertainty relation~\eqref{eq:UR_2New}.}


In what follows, the inner product on \GEN{$\CA^d$, denoted by $\braket{\,\cdot\,}{\,\cdot\,}$, is understood to be anti-linear in the first argument, and linear in the second.
Let $\omega$ be a strictly positive matrix on $\CA^d$ with eigenvalues arranged in ascending order:
\[
0 < \omega_1 \leq \omega_2 \leq \cdots \leq \omega_d.
\]
Let $X$ be an arbitrary complex matrix and $Y$ be a normal matrix. Then, $Y$ admits an eigenvalue decomposition of the form 
\[
Y = \sum_i y_i \ketbra{\phi_i}{\phi_i},
\] 
where $\{\phi_i \}$ is an orthonormal basis of eigenvectors. 
To prove the inequality \eqref{BW-new}, we first establish the following lemma.} \GRN{In what follows, we use the notation
\[
\omega_{vu} \defeq \braket{v}{\omega u}
\]
for $v,u\in\CA^d$. Since $\omega>0$, this defines an inner product on $\CA^d$.
}
\begin{Lemma}\label{lemma:A.1}
Let $\phi,\psi \in \CA^d$ be unit vectors with $\phi\perp \psi$,i.e., $\langle \phi|\psi\rangle=0$.  

{\rm (1)} The following inequality holds:
\begin{equation}\label{eq:A.1}
\frac{\omega_1+\omega_2}{\omega_1\omega_2}
\ge
\frac{\omega_{\phi\phi}+\omega_{\psi \psi}}{\omega_{\phi\phi}\omega_{\psi\psi} - |\omega_{\phi\psi}|^2}. 
\end{equation}

{\rm (2)} The inequality $ \|Y\|_\omega^2 \ge  \braket{\phi}{ Y \omega Y^\dagger \phi} +  \braket{\psi}{ Y \omega Y^\dagger \psi}$ holds. 
In particular, when $\phi = \phi_i$, the eigenvector of $Y$ corresponding to the eigenvalue $y_i$, 
\begin{equation}\label{eq:A.2}
\|Y\|_\omega^2 \ge  |y_i|^2 \omega_{\phi\phi} +  \braket{Y^\dagger \psi}{\omega Y^\dagger \psi}.
\end{equation}

{\rm (3)} Let $\xi = \alpha \phi +\beta \psi$ $(\alpha, \beta \in \CA)$. Then, 
\begin{equation}\label{eq:A.3}
\braket{\xi}{\omega \xi}
\ge |\beta|^2 \left( \frac{ \omega_{\phi\phi} \omega_{\psi\psi} - |\omega_{\phi\psi}|^2}{\omega_{\phi\phi}} \right).
\end{equation}

\end{Lemma}

{\it Proof.}
(1) Let $V = {\rm span}\{\phi,\psi
\}\subset {\mathbb C}^d$ and $P_V$ be the projection onto $V$. Then, $P_V \omega P_V$ is considered as a $2 \times 2$ matrix. 
Let $\sigma_1\leq\sigma_2$ denote the eigenvalues of $P_V\omega P_V$.
The Ky Fan's minimum principle (see e.g. \cite{BhatiaMatrixAnalysis}) says that for any $k=1,\ldots,d$,
\[
\sum_{i=1}^k \omega_i = {\rm min} \sum_{i=1}^k \langle v_i | \omega v_i \rangle,
\]
where the minimum is taken over all choices of orthonormal $k$-tuples $(v_1, \ldots, v_k)$ in ${\mathbb C}^d$. Applying this to $k=1$ and $k=2$, we obtain $0 < \omega_1 \le \sigma_1, \sigma_2$ and $\omega_1+\omega_2 \le \sigma_1 + \sigma_2$. 
The numerator and denominator in the right-hand side of \eqref{eq:A.1} are $\Tr(P_V \omega P_V)$ and the determinant of $P_V\omega P_V$, respectively. Therefore, 
\[
\frac{\omega_{\phi\phi}+\omega_{\psi\psi}}{\omega_{\phi\phi}\omega_{\psi\psi} - |\omega_{\phi\psi}|^2} =\frac{\sigma_1 + \sigma_2}{\sigma_1 \sigma_2} = \frac{1}{\sigma_1}+\frac{1}{\sigma_2}. 
\]
\GRN{Note here that, by the Schwarz inequality for the inner product $(v,u)\mapsto\omega_{vu}$, and since \(\phi\) and \(\psi\) are linearly independent, we have $\omega_{\phi\phi}\omega_{\psi\psi} - |\omega_{\phi\psi}|^2 >0$.}
\RED{
Since \(\sigma_1,\sigma_2\geq \omega_1\), we have
\[
\sigma_1\in
[\omega_1,\sigma_1+\sigma_2-\omega_1].
\]
For fixed \(\sigma_1+\sigma_2\), it is elementary to see that the function
\[
f(t):= \frac{1}{t}+\frac{1}{\sigma_1+\sigma_2-t}
\]
is strictly convex on the interval $[\omega_1,\sigma_1+\sigma_2-\omega_1]$.
Hence its maximum is attained at one of the endpoints. Since
\[
f(\omega_1)=f(\sigma_1+\sigma_2-\omega_1)
=
\frac{1}{\omega_1}
+
\frac{1}{\sigma_1+\sigma_2-\omega_1},
\]
we obtain
\[
f(\sigma_1)
=
\frac{1}{\sigma_1}+\frac{1}{\sigma_2}
\leq
\frac{1}{\omega_1}
+
\frac{1}{\sigma_1+\sigma_2-\omega_1}.
\]
Since \(\sigma_1+\sigma_2\geq \omega_1+\omega_2\), we have $\sigma_1+\sigma_2-\omega_1\geq \omega_2$, and hence we conclude
\[
\frac{\sigma_1+\sigma_2}{\sigma_1\sigma_2}
\leq
\frac{1}{\omega_1}+\frac{1}{\omega_2}
=
\frac{\omega_1+\omega_2}{\omega_1\omega_2}.
\]
}

(2)  \GEN{By definition of the $\omega$-norm, we have $\|Y\|_\omega^2 := \Tr Y^\dagger Y \omega
= \Tr Y\omega Y^\dagger$. 
Evaluating the trace in an orthonormal basis containing $\phi$ and $\psi$, and using the positivity of $Y\omega Y^\dagger$, we obtain
\[
\|Y\|_\omega^2
\geq
\braket{\phi}{Y \omega Y^\dagger \phi}
+
\braket{\psi}{Y \omega Y^\dagger \psi}.
\]
\RED{Applying this with $\phi=\phi_i$, and using the normality of $Y$, we have
$Y^\dagger\phi_i=\overline{y_i}\phi_i$. Therefore
\[
\braket{\phi_i}{Y\omega Y^\dagger\phi_i}
=
|y_i|^2\omega_{\phi_i\phi_i},
\qquad
\braket{\psi}{Y\omega Y^\dagger\psi}
=
\braket{Y^\dagger\psi}{\omega Y^\dagger\psi},
\]
which gives the assertion.}
}

(3) \GEN{We observe that
\begin{align*}
\braket{\xi}{\omega \xi}
&=
|\alpha|^2 \omega_{\phi\phi}
+
2{\rm Re}\bar{\alpha}\beta \omega_{\phi\psi}
+
|\beta|^2 \omega_{\psi\psi} \\
&=
\omega_{\phi\phi}
\left|
\alpha + \frac{\beta\omega_{\phi\psi}}{\omega_{\phi\phi}}
\right|^2
+
|\beta|^2
\left(
\omega_{\psi\psi}
-
\frac{|\omega_{\phi\psi}|^2}{\omega_{\phi\phi}}
\right) \geq
|\beta|^2
\left(
\frac{\omega_{\phi\phi}\omega_{\psi\psi}-|\omega_{\phi\psi}|^2}{\omega_{\phi\phi}}
\right).
\end{align*}}
\qed

We are now ready to prove the inequality \eqref{BW-new}. 
\begin{Theorem}\label{thm1}
For any complex matrix $X$, any normal matrix $Y$, and any strictly positive matrix $\omega$ on $\mathbb{C}^d$, the following inequality holds:
\[
\frac{\omega_1 + \omega_2}{\omega_1 \omega_2} \|X\|^2_\omega \|Y\|^2_\omega \geq \|[X, Y]\|^2_\omega 
\]
where $\omega_1$ and $\omega_2$ are the smallest and second-smallest eigenvalues of $\omega$, respectively. 
The bound is tight, in the sense that the coefficient $\frac{\omega_1+\omega_2}{\omega_1\omega_2}$ is the smallest possible constant for which the inequality holds for all complex matrices $X$ and all normal matrices $Y$.
\end{Theorem}

{\it Proof}. Using the eigenvalue decomposition of the form $Y = \sum_i y_i \ketbra{\phi_i}{\phi_i}$,
\[
[X,Y] = \Bigl(\sum_i \ketbra{\phi_i}{\phi_i}\Bigr) X Y -  \Bigl(\sum_i y_i \ketbra{\phi_i}{\phi_i}\Bigr)X =\sum_i X_i Y_i  
\]
where $X_i:= \ketbra{\phi_i}{\phi_i} X$ and $Y_i := Y - y_i \I$. Since $X^\dagger_i X_j =  \delta_{ij}X^\dagger \ketbra{\phi_i}{\phi_i} X$, we have 
\[
\|[X,Y]\|^2_\omega  = \Tr \Bigl(\sum_i X_i Y_i \Bigr)^\dagger \Bigl(\sum_j X_j Y_j \Bigr) \omega = \sum_i \braket{\xi_i}{ Y_i \omega Y^\dagger_i\ \xi_i}
\]
where $\xi_i := X^\dagger \phi_i$. 
On the other hand, 
\[
\|X\|_\omega^2 = \Tr(X^\dagger X \omega) = \Tr(X \omega X^\dagger)  = \sum_i \braket{\phi_i}{X \omega X^\dagger \phi_i} = \sum_i \braket{\xi_i}{ \omega \xi_i},
\]
and therefore,
\begin{equation}\label{eq:InSum}
\frac{\omega_1+\omega_2}{\omega_1\omega_2} \|X\|_\omega^2 \|Y\|_\omega^2 - \|[X,Y]\|_\omega^2 = \sum_i \braket{\xi_i}{ \left( \frac{\omega_1+\omega_2}{\omega_1\omega_2} \|Y\|^2_\omega  \omega -  Y_i \omega Y^\dagger_i \right) \xi_i}.  
\end{equation}
We show that each term on the right-hand side is nonnegative.

\GEN{For each $i$, decompose $\xi_i$ into its component along $\phi_i$ and its component orthogonal to $\phi_i$: $\xi_i=\alpha_i\phi_i+\beta_i \psi_i$, where $\psi_i\in\CA^d$ is a unit vector satisfying $\phi_i\perp \psi_i$, and $\alpha_i,\beta_i\in\CA$. 
Using Lemma \ref{lemma:A.1} by putting $\phi = \phi_i, \psi=\psi_i, \alpha = \alpha_i, \beta = \beta_i$, we obtain
\begin{align}
&\frac{\omega_1+\omega_2}{\omega_1\omega_2}\|Y\|_\omega^2 \braket{ \xi_i}{ \omega \xi_i} \nonumber \\
&\ge 
\frac{\omega_{\phi_i\phi_i}+\omega_{\psi_i\psi_i}}{\omega_{\phi_i\phi_i}\omega_{\psi_i\psi_i} - |\omega_{\phi_i\psi_i}|^2}
\left(|y_i|^2 \omega_{\phi_i\phi_i} + \braket{ Y^\dagger \psi_i}{ \omega Y^\dagger \psi_i}\right)
 |\beta_i|^2\frac{\omega_{\phi_i\phi_i}\omega_{\psi_i\psi_i}-|\omega_{\phi_i\psi_i}|^2}{\omega_{\phi_i\phi_i}} \label{eq:A.2new}\\
 &=\frac{\omega_{\phi_i\phi_i}+\omega_{\psi_i\psi_i}}{\omega_{\phi_i\phi_i}}
\left(|y_i|^2 \omega_{\phi_i\phi_i} + \braket{ Y^\dagger \psi_i}{ \omega Y^\dagger \psi_i}\right)
 |\beta_i|^2\nonumber 
\end{align}
\RED{
By $Y^\dagger\phi_i=\overline{y_i}\phi_i$, we have $Y_i^\dagger\phi_i=0$.
}
Therefore, $Y_i^\dagger \xi_i=\beta_i Y_i^\dagger \psi_i = \beta_i (Y^\dagger \psi_i-\overline{y_i}\psi_i)$, and we get  
\begin{align*}
\braket{\xi_i}{Y_i \omega Y_i^\dagger \xi_i}=
|\beta_i|^2
\left(
|y_i|^2\omega_{\psi_i\psi_i}
-
2{\rm Re}\left(
y_i\braket{\psi_i}{\omega Y^\dagger \psi_i}
\right)
+
\braket{Y^\dagger \psi_i}{\omega Y^\dagger \psi_i}
\right).
\end{align*}
It follows from these estimates that}
\begin{align}
&\braket{\xi_i}{ \left( \frac{\omega_1+\omega_2}{\omega_1\omega_2} \|Y\|^2_\omega  \omega -  Y_i \omega Y^\dagger_i \right) \xi_i} \nonumber\\
&\ge 
\frac{\omega_{\phi_i\phi_i}+\omega_{\psi_i\psi_i}}{\omega_{\phi_i\phi_i}}
\left(|y_i|^2 \omega_{\phi_i\phi_i} + \braket{ Y^\dagger \psi_i}{ \omega Y^\dagger \psi_i}\right)
 |\beta_i|^2 \\
& \quad  -|\beta_i|^2
\left(
|y_i|^2\omega_{\psi_i\psi_i}
-
2{\rm Re}\left(
y_i\braket{\psi_i}{\omega Y^\dagger \psi_i}
\right)
+
\braket{Y^\dagger \psi_i}{\omega Y^\dagger \psi_i}
\right) \nonumber\\
&= |\beta_i|^2 \left( |y_i|^2 \omega_{\phi_i\phi_i} + 2{\rm Re} y_i \braket{ \psi_i}{\omega Y^\dagger \psi_i} +
\frac{\omega_{\psi_i\psi_i}}{\omega_{\phi_i\phi_i}} \braket{ Y^\dagger \psi_i}{\omega Y^\dagger \psi_i}\right) \nonumber \\
&= |\beta_i|^2 \left(|y_i|^2 \frac{\omega_{\phi_i\phi_i}}{\omega_{\psi_i\psi_i}}\braket{ \psi_i}{ \omega \psi_i} + 2{\rm Re} y_i \braket{ \psi_i}{\omega Y^\dagger \psi_i} +
\frac{\omega_{\psi_i\psi_i}}{\omega_{\phi_i\phi_i}}\braket{ Y^\dagger \psi_i}{\omega Y^\dagger \psi_i} \right) \nonumber \\
&=|\beta_i|^2 \braket{\sqrt{\frac{\omega_{\phi_i\phi_i}}{\omega_{\psi_i\psi_i}}} \bar{y}_i \psi_i + \sqrt{\frac{\omega_{\psi_i\psi_i}}{\omega_{\phi_i\phi_i}}} Y^\dagger \psi_i}{
\omega \left( \sqrt{\frac{\omega_{\phi_i\phi_i}}{\omega_{\psi_i\psi_i}}} \bar{y}_i \psi_i + \sqrt{\frac{\omega_{\psi_i\psi_i}}{\omega_{\phi_i\phi_i}}} Y^\dagger \psi_i  \right)}  \ge 0 \label{eq:A.5}
\end{align}
which completes the proof of the inequality stated in Theorem~\ref{thm1}. 

\GRN{
It remains to show that the coefficient is tight.
For this purpose, it suffices to exhibit matrices $X$ and $Y$ for which equality is attained and the two sides of the inequality are nonzero.
A simple Hermitian example is given by the same construction as in~\eqref{eq:Optimal}:
\[
X:=\omega_2\ketbra{1}{1}-\omega_1\ketbra{2}{2},
\qquad
Y:=\ketbra{1}{2}+\ketbra{2}{1},
\]
where $|1\rangle$ and $|2\rangle$ are orthonormal eigenvectors of $\omega$ corresponding to its smallest and second-smallest eigenvalues $\omega_1$ and $\omega_2$, respectively.

A direct computation gives
\[
\|X\|_\omega^2
=
\omega_1\omega_2(\omega_1+\omega_2),
\qquad
\|Y\|_\omega^2
=
\omega_1+\omega_2.
\]
Moreover,
\[
[X,Y]
=
(\omega_1+\omega_2)
\left(
\ketbra{1}{2}-\ketbra{2}{1}
\right),
\]
and hence
\[
\|[X,Y]\|_\omega^2
=
(\omega_1+\omega_2)^3.
\]
Therefore,
\[
\frac{\omega_1+\omega_2}{\omega_1\omega_2}
\|X\|_\omega^2\|Y\|_\omega^2
=
\frac{\omega_1+\omega_2}{\omega_1\omega_2}
\omega_1\omega_2(\omega_1+\omega_2)^2
=
(\omega_1+\omega_2)^3
=
\|[X,Y]\|_\omega^2.
\]
Thus equality is attained for nonzero Hermitian matrices $X$ and $Y$.
Consequently, the coefficient
$\frac{\omega_1+\omega_2}{\omega_1\omega_2}$ cannot be decreased in general.
This proves the tightness of the bound.
\qed
}

\section{Key Lemma for the Proof of the Uncertainty Relation \eqref{eq:UR_2New}}

The proof of the uncertainty relation \eqref{eq:UR_2New} is based on the following lemma: 
\begin{Lemma}\label{lem:stronger}
Let $\omega$ be a strictly positive matrix and let $c>0$ be a constant, possibly depending on $\omega$. Suppose that
\begin{equation}\label{eq:assume2}
c \|X\|_\omega^2 \|Y\|_\omega^2
	\geq \|[X, Y]\|_\omega^2
\end{equation}
holds for any complex matrix $X$ and any normal matrix $Y$. 
Then the stronger relation
\begin{equation}\label{eq:strongURapp}
c \left( \|X\|_\omega^2 \, \|Y\|_\omega^2 - |\Tr(X^\dagger Y \omega)|^2 \right)
	\geq \|[X, Y]\|_\omega^2
\end{equation}
also holds.
\end{Lemma}

{\it Proof.} 
The Schwarz inequality for the inner product $\langle X|Y \rangle_\omega := \Tr(X^\dagger Y \omega)$ implies that if $\|Y\|_\omega = 0$, then $\Tr(X^\dagger Y \omega) = 0$. Therefore, in the case $\|Y\|_\omega = 0$, the statement holds trivially: $0 \geq 0$.

We henceforth assume $\|Y\|_\omega \neq 0$. 
In view of the identity
\begin{align}
[X, Y] = [X + tY, Y]
\end{align}
for any $t \in \mathbb{C}$, the assumption \eqref{eq:assume2} implies
\begin{equation}\label{eq:B.2}
c \|X + tY\|_\omega^2 \|Y\|_\omega^2 \geq \|[X, Y]\|_\omega^2.
\end{equation}
Letting $t = -\frac{\overline{\Tr (X^\dagger Y \omega)}}{\|Y\|_\omega^2}$, we obtain
\begin{align*}
\|X+tY\|_\omega^2 &= 
\|X\|_\omega^2 + 2 {\rm Re} \left( t \Tr(X^\dagger Y\omega) \right)+ |t|^2 \|Y\|_\omega^2 \\
&=\|X\|_\omega^2  - \frac{|\Tr (X^\dagger Y\omega)|^2}{\|Y\|_\omega^2}.
\end{align*}
Substituting this into \eqref{eq:B.2} yields \eqref{eq:strongURapp}.
\hfill \qed 

\begin{Remark}
\GEN{
To derive the uncertainty relation~\eqref{eq:FamUR}, it would have been sufficient to prove Theorem~\ref{thm1} only for Hermitian matrices $X$ and $Y$.
However, deriving the uncertainty relation~\eqref{eq:UR_2New} requires applying the inequality of Theorem~\ref{thm1} via this lemma. This is why Theorem~\ref{thm1} had to be formulated with $X$ allowed to be an arbitrary complex matrix. 
Indeed, in the proof above, the original inequality is applied to $X+tY
$ for an arbitrary complex number $t$.
}
\end{Remark}

\section{Details of Qubit Calculations}
Here, we provide all calculations for qubit case in detail. In what follows, we use the same notation for state, and observables as in equations \eqref{Bloch} and \eqref{AB} in the main text: 
\[
\rho = \frac{1}{2}(\mathbb{I} + \bm{r} \cdot \bm{\sigma}) := \frac{1}{2}(\mathbb{I} + \sum_{i=1}^3 r_i \sigma_i)
\]
and 
\[
A=a_0\I+\bm a\cdot\bm \sigma, \quad B=b_0\I+\bm b\cdot\bm \sigma,
\]
where $\bm{r} = (r_1, r_2, r_3) \in \mathbb{R}^3$ is a Bloch vector of $\rho$, $\bm a=(a_1,a_2,a_3), \bm b=(b_1,b_2,b_3) \in \mathbb{R}^{3}$ are the ``spin" direction of $A$ and $B$ with $a_0, b_0 \in \mathbb{R}$, and $\bm{\sigma} = (\sigma_1, \sigma_2, \sigma_3) = (\sigma_x, \sigma_y, \sigma_z)$ denotes the vector of Pauli matrices. 
We denote the Euclidean inner product, Euclidean norm, and cross product by
$\bm{a} \cdot \bm{b} := \sum_{i=1}^3 a_i b_i$, $|\bm{a}| := \sqrt{\sum_{i=1}^3 a_i^2}$, and $\bm{a} \times \bm{b} := (a_2 b_3 - a_3 b_2,\, a_3 b_1 - a_1 b_3,\, a_1 b_2 - a_2 b_1) = (\sum_{j,k=1}^3 \epsilon_{ijk} a_j b_k)_{i=1}^3$ with $\epsilon_{ijk}$ being the Levi-Civita symbol, respectively. 
In what follows, we use the following formulae:
\[
(\bm{a} \cdot \bm{\sigma})(\bm{b} \cdot \bm{\sigma})
=(\bm{a} \cdot \bm{b}) \I +i(\bm{a} \times \bm{b})\cdot\bm{\sigma}, \qquad
(\bm{a} \cdot \bm{\sigma})^2
=|\bm{a}|^2 \I, \qquad
\Tr(\rho(\bm{a} \cdot \bm{\sigma}))=\bm{a} \cdot \bm{r}.
\]

\begin{Theorem}
The exact uncertainty relation \eqref{eq:UR_3}
\[
V_\rho(A)\, V_\rho(B) = 
\frac{1}{4} \left| \langle [A,B] \rangle_\rho \right|^2 + 
{\rm Cov}_\rho(A, B)^2 + 
\frac{1 - P(\rho)}{4} \| [A, B] \|^2
\]
holds for all two-level systems.
\end{Theorem}
{\it Proof.} 
Using $\Tr(\bm{a} \cdot \bm{\sigma})=0$, direct computation shows that
\begin{align*}
\langle A^2 \rangle_\rho &= \Tr(\rho (a_0^2\I  + 2a_0 (\bm{a} \cdot \bm{\sigma}) + |\bm{a}|^2\I )) \nonumber \\
&= a_0^2 + 2a_0 (\bm{a} \cdot \bm{r}) + |\bm{a}|^2, \\
\langle A \rangle_\rho &= \Tr(\rho(a_0\I + \bm{a} \cdot \bm{\sigma})) = a_0 + \bm{a} \cdot \bm{r}. 
\end{align*}
Accordingly, the variance reads
\[
V_\rho(A)= \langle A^2\rangle_\rho-\langle A\rangle_\rho^2 = |\bm{a}|^2-(\bm{a}\cdot\bm{r})^2, \ V_\rho(B)= |\bm{b}|^2-(\bm{b}\cdot\bm{r})^2.
\]
The commutator and the anticommutator are given by
\[
[A,B] =2i (\bm{a}\times\bm{b})\cdot\boldsymbol{\sigma},
\]
\[
\{A,B\} =2a_0b_0 \I + 2a_0(\boldsymbol{b}\cdot\boldsymbol{\sigma}) + 2b_0(\boldsymbol{a}\cdot\boldsymbol{\sigma}) + 2(\boldsymbol{a}\cdot\boldsymbol{b})\I.
\]
One obtains the first two expressions \eqref{bounds} as follows: 
\begin{align}
\frac{1}{4}|\langle[A,B]\rangle_\rho|^2 &= \frac{1}{4}|2i\Tr(\rho((\bm{a}\times\bm{b})\cdot\boldsymbol{\sigma}))|^2 = |(\bm{a}\times\bm{b})\cdot\bm{r}|^2,\nonumber\\[2mm]
{\rm Cov}_\rho(A,B)^2 &= \left|\frac{1}{2}\langle\{A,B\}\rangle_\rho-\langle A\rangle_\rho\langle B\rangle_\rho\right|^2  \nonumber \\
&=|a_0b_0+a_0(\bm{b}\cdot\bm{r})+b_0(\bm{a}\cdot\bm{r})+\bm{a}\cdot\bm{b} -(a_0 +\bm{a}\cdot\bm{r})(b_0+\bm{b}\cdot\bm{r})|^2  \nonumber \\
&=|\boldsymbol{a}\cdot\boldsymbol{b}-(\boldsymbol{a}\cdot\bm{r})(\boldsymbol{b}\cdot\bm{r})|^2.\nonumber
\end{align}
Since the square of the norm of the commutator and the purity is calculated as 
\[
\|[A,B]\|^2 = \|2i(\bm{a}\times\bm{b})\cdot\bm{\sigma}\|^2
=8|\bm{a}\times\bm{b}|^2,
\]
\[
P(\rho)=\lambda_1^2+\lambda_2^2 = 1-2\lambda_1\lambda_2 = 1- \frac{1-|\bm{r}|^2}{2}=\frac{1+|\bm{r}|^2}{2},
\]
the third equation \eqref{eq:thirdeq} follows from
\[
\frac{1 -P(\rho)}{4} =  \frac{1-|{\bm r}|^2}{8}.
\]
The equality \eqref{iesNFN} follows from 
\begin{align}
\|[A,B]\|^2_\rho 
=\|2i(\bm{a}\times\bm{b})\cdot\bm{\sigma}\|^2_\rho
=4\Tr(\rho(|\bm{a}\times\bm{b}|^2\I)) = 4 |\bm{a}\times\bm{b}|^2
=\frac{1}{2}\|[A,B]\|^2. \nonumber
\end{align}
Combining these equations, we conclude \eqref{eq:UR_3}. \qed

\begin{Proposition}
The expression \eqref{Zheng} coincides with our bound \eqref{Bdd2lev}:
\[
\frac{1-P(\rho)}{8}(\xi(A,A)\xi(B,B)-\xi(A,B)^2) =\frac{\lambda_1\lambda_2}{\lambda_1+\lambda_2}\|A,B\|^2_\rho.
\]
\end{Proposition}

{\it Proof.}
By the definition of $\xi(A,B)$, we can calculate as
\[
\xi(A,B) = 2\Tr(AB) -\Tr(A)\Tr(B)
= 2(2a_0b_0 +2\bm{a}\cdot\bm{b})-4a_0b_0 = 4\bm{a}\cdot\bm{b}.
\]
Therefore, we obtain
\begin{align*}
&\frac{1-P(\rho)}{8}(\xi(A,A)\xi(B,B)-\xi(A,B)^2) = 2(1-P(\rho))(|\bm{a}|^2 |\bm{b}|^2 - (\bm{a}\cdot\bm{b})^2)  \\ 
&= 2(1-P(\rho))|\bm{a}\times\bm{b}|^2= \frac{(1-P(\rho))\|[A,B]\|^2}{4} \\ 
&=\frac{\lambda_1\lambda_2}{\lambda_1+\lambda_2}\|A,B\|^2_\rho
\end{align*}
which shows the assertion. \qed

\bibliographystyle{unsrt} 
\bibliography{ref_UR}     

\end{document}